\def\WantIEEETran{0} 
\ifodd\WantIEEETran
\documentclass[10pt, conference, x11names]{IEEEtran}
\else
\documentclass[10pt, conference, x11names]{IEEEconf}
\fi
\usepackage{listings}
\usepackage{amsmath}
\usepackage{orcidlink}
\usepackage{amssymb}
\usepackage{balance}
\usepackage{hyperref}
\def\nobreakseq{\nobreak\hskip0pt\hbox}
\title{Advances in Semantic Patching for HPC-oriented Refactorings with Coccinelle}
\ifodd\WantIEEETran
\author{
\IEEEauthorblockN{Michele MARTONE\IEEEauthorrefmark{1}
$^{\orcidlink{0000-0003-3239-8554}[0000-0003-3239-8554]}$
and
Julia LAWALL\IEEEauthorrefmark{2}
$^{\orcidlink{0000-0002-1684-1264}[0000-0002-1684-1264]}$
}
\IEEEauthorblockA{\IEEEauthorrefmark{1}
Leibniz Supercomputing Centre,
Garching bei Muenchen, Germany\ \ \ \ 
Email: michele.martone@lrz.de\\
}
\IEEEauthorblockA{\IEEEauthorrefmark{2} Inria, Paris, France
\ \ \ \ 
Email: julia.lawall@inria.fr\\
}
}
\else
\author{
Michele MARTONE
\begin{affiliation}
Leibniz Supercomputing Centre, Garching near Munich, Germany
\end{affiliation}
\and
Julia LAWALL
\begin{affiliation}
Inria, Paris, France
\end{affiliation}
}
\fi
\def\OPENMP{\textsc{Open\-MP}}
\def\OPENACC{\textsc{Open\-ACC}}
\usepackage{color}
\makeindex
\usepackage{graphicx}
\usepackage{xcolor}
\def\difffgcolor{Snow1}
\def\coccifgcolor{Ivory2}
\def\coccihcolor{olive}
\def\coccipluscolor{blue}
\def\cocciminuscolor{red}
\def\coccimcolor{violet}
\def\coccipcolor{darkgray!85}
\def\coccigcolor{darkgray!75}
\def\coccibgcolor{Ivory1}
\lstdefinelanguage{diff}{
frame=none,
morecomment=[l][\color{red}]TODO,
morecomment=[l][\color{red}]FIXME,
morecomment=[f][\color{cyan}][0]\#,
morecomment=[f][\color{red}][0]\-,%
morecomment=[f][\color{blue}][0]+,%
morecomment=[f][\color{Aquamarine4!70!black}][0]@,
	backgroundcolor=\color{\difffgcolor},
}
\lstdefinelanguage{myC}{
	language=C,
	basicstyle={\ttfamily},
	morecomment=[f][\color{blue}][0]\#
}
\lstdefinelanguage{cocci}{
	language=bash,
	basicstyle={\ttfamily\small
	\color{\coccipcolor}
	},
	frame=none,
	morekeywords={comments,type,symbol,identifier,expression,name,attribute,idexpression,statement,position,when,field,forall,format,fresh,exists,strict,any,parameter,list,initializer,constant,virtual,typedef,declarer,iterator,local,global,declaration,TODO,operator,binary,assignment,pragmainfo},
	keywordstyle={\bfseries\color{violet}},
	showstringspaces=false,
morecomment=[l][\color{cyan}]//,
morecomment=[f][\color{purple}][0]cocci,
morecomment=[f][\color{purple}][0]print ,
morecomment=[f][\color{\coccimcolor}][0]{\&},
morecomment=[f][\color{\coccimcolor}][0]{\|},
morecomment=[f][\color{\coccipluscolor}][2]{match},
morecomment=[f][\color{purple}][2]{cocci},
morecomment=[f][\color{purple}][4]{(},
morecomment=[f][\color{purple}][0]coccinelle,
morecomment=[f][\color{\coccihcolor}][0]@,
morestring=[d]",
morecomment=[f][\color{\cocciminuscolor}][0]\-,
morecomment=[f][\color{\coccipluscolor}][0]+,
	backgroundcolor=\color{\coccibgcolor},
}
\def\mycoccisnip#1{\lstinline[columns=fixed,language=cocci]{#1}}
\def\mycsnip#1{\lstinline[columns=fixed,language=myC]{#1}}
\def\mycpragmasnip#1{\fcolorbox{\coccifgcolor}{\coccibgcolor}{{\color{blue}\texttt{#1}}}}
\lstdefinelanguage{coccioutput}{
	language=bash,
	basicstyle={\ttfamily
	\color{darkgray}
	},
	frame=none,
	keywords={spatch},
	morekeywords={},
	keywordstyle={\bfseries\color{blue}},
	showstringspaces=false,
	morecomment=[f][\color{black}][0]+,
	backgroundcolor=\color{white},
	backgroundcolor=\color{white},
}
\def\coccipcode#1{\fcolorbox{\coccifgcolor}{\coccibgcolor}{{\color{\coccipluscolor}\texttt{#1}}}} 
 
\def\coccipelem#1{\fcolorbox{\coccifgcolor}{\coccibgcolor}{{\color{\coccipcolor}\texttt{#1}}}} 
\def\coccimelem#1{\fcolorbox{\coccifgcolor}{\coccibgcolor}{{\color{\coccimcolor}\texttt{#1}}}} 

\def\coccihelem#1{\fcolorbox{\coccifgcolor}{\coccibgcolor}{{\color{\coccihcolor}\texttt{#1}}}}

\newcommand{\para}[1]{\ifodd\WantIEEETran \paragraph*{\textbf{#1}} \else \paragraph*{#1.} \fi}

\usepackage{ifthen}
\newboolean{showcomments}
\setboolean{showcomments}{true}
\ifthenelse{\boolean{showcomments}}
{ \newcommand{\mynote}[3]{\textcolor{#3}{
    \fbox{\bfseries\sffamily\scriptsize#1}
    {\small$\blacktriangleright$\textsf{\emph{#2}}$\blacktriangleleft$}}}}
{ \newcommand{\mynote}[2]{}}

\newcommand{\hidefornow}[1]{}

\begin{document}
\maketitle
\begin{abstract}
Currently, the most energy-efficient hardware platforms for floating point-intensive calculations (also known as High Performance Computing, or HPC) are graphical processing units (GPUs).
However, porting existing scientific codes to GPUs can be far from trivial.
This article summarizes our recent advances in enabling
machine-assisted, HPC-oriented refactorings with reference to existing
APIs and programming idioms available in C and C++.
The tool we are extending and using for the purpose is called Coccinelle.
An important workflow we aim to support is 
that of writing and maintaining tersely
written application code, while deferring circumstantial, ad-hoc,
performance-related changes to specific, separate
rules called ``\textit{semantic patches}''.
GPUs currently offer very limited debugging facilities.
The approach we are developing aims at preserving
intelligibility, longevity, and relatedly, debuggability of
existing code on CPUs, while at the same time enabling HPC-oriented code
evolutions such as introducing support for GPUs, in a scriptable and possibly parametric manner.
This article sketches a number of self-contained use cases, including further HPC-oriented cases which are independent from GPUs.
\end{abstract}
\section{Introduction}

In today's computing environment, the best hardware on which to execute
demanding HPC applications, in terms of both execution time and energy
costs, is often the GPU, rather than the CPU.  Nevertheless, much
HPC software was written in the CPU era.  Continuing to use
these valuable applications effectively thus requires transforming their source code to
use GPU libraries.  Such transformations must be done systematically.  The
large size and complexity of the codebases involved suggest that it is desirable
to do as much as possible of these transformations automatically.

In the Linux kernel developer community, the tool Coccinelle\footnote{\url{https://github.com/coccinelle/coccinelle}} is regularly
used to perform large scale transformations to the Linux kernel code base.
Coccinelle targets transformation of C code and has been designed around
the notion of a {\em semantic patch}.  A semantic patch is a change
specification written in terms of lines to add and remove, like a
traditional patch, but is applied taking into account the type system and
the control flow of the target programming language.  Beyond the Linux
kernel, Coccinelle is also used by developers of some other systems
software such as
\textsc{systemd}\footnote{\url{https://github.com/systemd/systemd/tree/main/coccinelle}}
and
\textsc{Zephyr OS}\footnote{\url{https://github.com/zephyrproject-rtos/zephyr/tree/main/scripts/coccinelle}}.
Recently \cite{c3po}, Coccinelle has also been applied to the cosmological
software \textsc{Gadget}, with the goal of the targeted transformation of
specific data from the array of structures (AoS) to the structure of arrays
(SoA) representation.

This article is motivated by
a new training course \cite{martone_20250122_coccinelle_training}
on Coccinelle that has been offered at the
Leibniz Supercomputing Centre (LRZ), to facilitate the use of Coccinelle by HPC programmers.
This training contains new introductory didactic material relevant to HPC,
such as operations on floating-point numeric
arrays and loop constructs with parallel APIs, and includes an overview of possible applications.

\section{Literature Overview}
\label{sec:litov}
We now contextualize Coccinelle and the present work within related literature and techniques.

A \textit{Semantic patch}, written in the \emph{Semantic Patch Language}
(\textsc{SmPL}), as introduced by Coccinelle
\cite{lawall_muller_2018_ten_years_atc},
refers to a code-change specification that has a form similar to a \textsc{POSIX} \emph{unified code patch},
indicating removed and added lines of code.
Instead of matching text only, as for a \textsc{POSIX} unified code patch,
a semantic patch is matched
against a program's abstract syntax tree (AST) and control-flow graph
(CFG).  Furthermore, change specifications are abstracted by {\em
metavariables} making it possible for subterms to match arbitrary subterms
such as types, expressions, and statements.  The use of metavariables makes
a semantic patch generic, allowing a single change specification to be
applied across a code base.  Matching against the AST, rather than
the source code text as done for a \textsc{POSIX}
patch, ensures that the language syntax is respected, including matching
complete variable names, rather than substrings, and respecting the
relative precedence of infix binary operators.  Finally, respecting the CFG
makes it possible for the matching of a
semantic patch to follow the control flow of execution, for example around
a loop, thus respecting the execution semantics.

Coccinelle was created to provide the \textsc{Linux} kernel developers with
a means to succinctly describe transformation rules and apply them.
Exchange of \textsc{POSIX} patches is a common working practice within the \textsc{Linux} kernel community,
and, by analogy, Coccinelle has become the de-facto tool to express complex refactorings within the \textsc{Linux} kernel community,
by means of files containing semantic patches.

Since the 1990s, the term \textit{``Refactoring''} \cite{Fowler1999} has
been used to denote semantics-preserving stepwise code manipulations that
introduce or enforce certain desirable qualities (e.g., absence of
\textit{code smells}) that would otherwise stem from best practices of design and implementation.
Existing IDEs (Integrated Development Environments)
often support a fixed number of specific \textit{refactoring}s that are expected to be
of general interest.
Many of these are quite simple.  However, as they are
implemented using a general-purpose language, such as C++, they can be
arbitrarily complex.
Nevertheless, as the set of refactorings offered by
an IDE is fixed, it cannot be easily adapted to the specific properties of a software
project.  Furthermore, some desirable transformations may not precisely
preserve the semantics, putting them outside of the scope of refactorings.
For example, a change introducing parallelism may
affect the result due to changing the order of arithmetic
operations; such changes can be tolerable but are not always desirable.
To counter such situations, Coccinelle allows fine-grained control of the set of code
fragments that are to be transformed.

The C \textit{preprocessor} (Sec.~6.10 in \cite{C23}) 
enables several techniques for selective compilation, 
code selection, and avoiding redundancy.
That works by introducing macros to abstract code from
semantics (e.g., masking loops within macro definitions),
and can involve a significant cost in clarity
and an increased risk of bugs,
especially when nesting several levels of macros.
For these reasons, despite supporting some changes in code semantics equivalent to the
\emph{HPC-oriented refactorings} we are interested in, 
preprocessor tricks can quickly reach their limits, and lead to problems.
Extensive usage of preprocessor macros is regarded as dangerous and is therefore discouraged.

The LLVM project offers a templated C++-based API that allows
direct and low-level matching and manipulation of the AST \cite{CRE},
and output in the original language (C or C++).
This allows matching and modifying potentially vast swathes of code
by following an imperative programming model. Such a framework can be used
to implement refactorings.
This solution requires matching constructs at the exact \textit{grammar production}
level, which in many cases may be unnecessarily picky,
in that it requires writing \textit{refactoring programs} that are more
verbose (and that require more effort) than is desirable.
Consequently, it is mostly authors of IDEs or underlying tools such as \texttt{clang-tidy} 
who use this API to prepare catalogues of refactorings.

The ROSE compiler infrastructure \cite{quinlan_2011_rose}\cite{rosewww} has 
a broad scope of applications that includes HPC-oriented source-to-source translation,
and has been created for the needs of the United States of America's Department of Energy.
Differently than \textsc{SmPL} for Coccinelle, and similarly to LLVM, ROSE offers a C++ interface to express refactorings.

Most recently, machine learning-backed \textit{code assistant chatbot}
services started to be offered by a few major corporations
(e.g., \textsc{GitHub Copilot} by {Microsoft}, \textsc{ChatGPT} or \textsc{CodeGPT} by \textsc{OpenAI}, etc.)
Such chatbots are operated by the user by means of natural language in written form.
They can generate code or manipulate existing code in a way requested by the user.
Their inner mechanisms (source code and training datasets) are often
somewhat opaque, just as \textit{explainability} of machine learning
techniques remains a challenge.
These approaches are substantially different from the one proposed here, in
that they offer full automation and little user control.  Our approach does
not involve any form of machine learning.

Several HPC libraries offer abstractions by creating a separation 
layer between an actual data structure (and where it resides) and the syntax used to
access it.
This is one of the features of e.g., \textsc{Kokkos} \cite{edwards_trott_sunderland_2014_kokkos_data_layout_change}, where
the user can make a compile-time choice between
\textit{Structure of Arrays (SoA)}-like and \textit{Array of Structures (AoS)}-like
accesses with a unified syntax.
Nevertheless, it is likely that most numeric expressions will require a rewrite to use \textsc{Kokkos} array access notation.

As an alternative to large-scale expression rewriting, modern C++ allows introduction of a number of classes equipped with
\textit{inline accessor member functions}, and custom
\textit{operator overloads}
in a way that AoS access syntax can be fully preserved across the code, and SoA accesses occur under the hood.
Provided compiler optimization is active (so as to cancel the underlying overhead),
such techniques ease achieving vectorizable arrays access, and restrict code changes mostly
to data structure definitions.
Compiling the sources without any optimization and with debug flags enabled will
mostly likely retain object code at each inline function call, in places where
optimization flags would have led to a mere memory location access.
This consequence of the inherently more complicated logic underneath
can slow down the debuggable code to a degree
where reaching the point of a crash can be problematic.

In our past work \cite{c3po}, we described a use case where we
performed an AoS $\rightarrow$\ SoA change explicitly on the entire
\textsc{Gadget} cosmological code \cite{springel2005cosmological}.
That intervention was recommended by colleagues
(see their pilot study in \cite{BaruffaIHK16})
in order to achieve a higher rate of compiler auto-vectorization \cite[Ch.~8]{Pennycook2015143}.
In addition to the data structure definitions (a mere few hundred lines of code, which we could have changed by hand), the
rules we created patched many tens of
array-accessing expressions within each of thousands of loops.
We left the original code to be straight C, and performed our changes by means of a collection of Coccinelle rules.
If using no optimization flags and/or using debug flags there is no penalty, because there is no syntax abstraction obscuring the semantics.

In this example, Coccinelle makes it possible to provide the domain
scientists with transformation rules that allow them
to continue developing the original AoS code, which can be a bit clearer to
understand than SoA code.
A \textit{refactored} SoA copy of the code can be obtained by invoking Coccinelle on it.
In such a workflow, one may want to use version control for the semantic patches only, which are much terser and less redundant than the transformed code.
The approach also allows easily fine-grained control of where the transformation is applied:
specified quantities can be kept in AoS form if this is desired for modularization or organizational reasons.
This style of working could allow developing multiple concurrent \textit{experimental derivations} of a codebase, without the burden of having to maintain several branches.

\section{Enabled HPC Refactorings}
\label{sec:enrefs}
Coccinelle has been recently enhanced with features that allow
expressing HPC-oriented refactorings that were previously not expressible.
Here is an overview.

\para{Interfacing with an instrumentation API}
	Several tools for instrumentation (e.g.,
	\textsc{LIKWID},\footnote{\url{https://github.com/RRZE-HPC/likwid}}
	\textsc{Score-P},\footnote{\url{https://score-p.org/}}
        \textsc{Caliper}.\footnote{\url{https://github.com/LLNL/Caliper}}
        etc.) offer an API (also referred as a \textit{marker API}) to collect code performance metrics at runtime.
	Introduction and removal of instrumentation syntax is one of the
        simplest possible tasks that we present.
	To achieve that, one would specify an \textsc{SmPL} pattern describing the code
        that should be measured by the marker API, for instance by enclosing it with
	calls to start or stop measurement collection.
	As an example, the following semantic patch introduces calls to the LIKWID macros.
\begin{lstlisting}[language=cocci]
@@ @@
  #include <omp.h>
+ #include <likwid-marker.h>

@@ @@
 #pragma omp ...
 {
+ LIKWID_MARKER_START(__func__);
  ...
+ LIKWID_MARKER_STOP(__func__);
 }
\end{lstlisting}
	This semantic patch uses two rules.
	The first one introduces the \textsc{LIKWID} header
	just after the \textcolor{blue}{\texttt{\#include <omp.h>}} line.
        The second rule locates (\textit{any}) \textcolor{blue}{\texttt{\#pragma omp}} line
        followed by a block within braces, and inserts \textit{instrumentation} to start/stop performance 
	collection, passing the compiler-provided \mycsnip{__func__} string
        as a parameter specifying the current function name as a computation phase label.
	In practice, one could refine the pattern to be more selective in
        choosing such code locations. For instance,
	restricting to specific numeric kernels, or around the specific
        areas of the code that one is concentrating on at a given time.
	The parametrical nature of the match specification allows flexibility
	and experimentation in deciding which areas of the code to instrument, perhaps only transitorily.

\para{\OPENMP's \mycsnip{declare variant}}
	Since version 5, the specification document of \OPENMP{} \cite{OMP6} introduces a notation
	for the definition and automatic selection of hardware-optimized functions, referred to as \textit{variants}.
	A \textit{variant} is only compiled if its specific \textit{istruction set architecture} (ISA) is selected at compile time.
	If not, only the source code corresponding to the base version is be compiled.
	This notation supports variants via
        one \textcolor{blue}{\tt{\#pragma omp declare variant ...}} line
        per variant just before the base function definition.
	Each such a clause specifies which function is to be treated as a variant.
	We illustrate here an approach to introducing variants on a large scale with the help of Coccinelle.

	Let us assume that we have a source file with an arbitrary number of functions that we wish to diversify in this way.
	For each candidate function (in the example here, we select those whose name matches the regular expression \texttt{"kernel"}),
	two clones are created (their names being derived from the base function name) and declared as variants
	by means of \textcolor{blue}{\tt{\#pragma omp declare variant}}
        clauses just above the base function.
	The function clones can be created easily by matching function definitions with the \textsc{SmPL} code
        \mycoccisnip{T f (PL) \{ SL \}} (line 13).
	Such an arrangement matches a previously defined \mbox{\mycoccisnip{type T}},
	a function name (\mycoccisnip{identifier}) \mycoccisnip{f}, and
	its \mycoccisnip{parameter list PL}.
	These components are then used to insert two function clones (lines 9 and 10 of the listing, prefixed by \coccipcode{+}).
	The clones' names are produced via \coccimelem{fresh identifier} (lines 6 and 7), extending (via the \mycoccisnip{##} operator) the original
        function name (as matched by \mycsnip{f}).
	The names of the clones of each original function are constructed by prefixing 
	respectively \mbox{\mycsnip{"avx512_"}} and \mbox{\mycsnip{"avx10_"}} to \mycsnip{f}, and the
	\mycpragmasnip{match} clause of each clone indicates to the compiler under which target architecture it shall be used.

\begin{lstlisting}[language=cocci]
@@
type T;
identifier f =~ "kernel";
parameter list PL;
statement list SL;
fresh identifier f512 = "avx512_"##f;
fresh identifier f10 = "avx10_"##f;
@@
+ T f512 (PL) { SL }
+ T f10 (PL) { SL }
+#pragma omp declare variant(v512_f) match(device={isa{"core-avx512"}})
+#pragma omp declare variant(v10_f)  match(device={isa{"core-avx10"}})
  T f (PL) { SL }
\end{lstlisting}

	At this stage, the two function clones corresponding to {\tt f512} and {\tt f10} are identical to the base function identified by \mycsnip{f}.
	To obtain the hardware-specific variants we seek, we would still have to write a few extra rules
	that enact specific transformations on them.
	Such rules could reside in this semantic patch and \textit{inherit} the \mycsnip{f512} and \mycsnip{f10} metavariables, or
	be completely independent and reside in a different semantic patch file, only to be
        applied in a separate step, exploiting the pragma and the specific function
        name (which suggests it is a clone) to use an architecture-specific rule.

	In either case, modifying the clones would involve rules introducing changes
	that are optimal for the respective architectures (here we alluded to either AVX-10 or AVX-512).
	Such rules would modify statements; e.g., introduce calls to \textit{intrinsics} or arrange for specific properties or types in variable definitions.
	Such rules must necessarily be very program-specific, and their
	discussion does not belong here.

\para{Function cloning and introduction of attributes for function multiversioning}
Independent of OpenMP, the GCC and CLANG compilers offer a C/C++ extension
to create a form of function variant, called
	\textit{function multiversioning},
	by means of specific \textit{attributes}.
	The first one (\texttt{target\_clones}) allows implicit creation of processor-specific copies 
	of a function \cite{GFM}, as well as automatic selection of these at runtime.
	The second one (\texttt{target}) requires multiple function definitions, each of which can also contain several custom compilation options.
	These extensions ease the production of binary libraries with object code
        optimized for different processors within the same base architecture family.
	Here in this example, the base function is being prefixed by 
	\mycsnip{__attribute__((target("default")))} 
	and variations of this function are marked as such, e.g.,
	\mycsnip{__attribute__((target("avx512")))}.
	To automate creation of such function clones with Coccinelle we proceed as we have shown for \textcolor{blue}{\texttt{\#pragma omp declare variant}} (previous listing).
	In a second step (following listing) we match on the attribute expression
	containing \textcolor{blue}{\texttt{"avx512"}};
	the rules which would follow this match (not shown here) would have to 
	to introduce specific specializations (e.g. usage of intrinsics).
	
\begin{small}
\begin{lstlisting}[language=cocci]
@@
identifier f;
type T;
@@
__attribute__((target(...,"avx512",...)))
T f(...)
{
+ // add and modify avx512-specific code only
  ...
}
\end{lstlisting}
\end{small}
	If matching is needed on compiler-specific attributes
	that are not prefixed by ``\mycsnip{__}'', Coccinelle would require
	them to be specified as such via \mycoccisnip{attribute name}.

\para{Bloat and clone removal}
	Given a project with a long development history, we may be interested in removing obsolete code.
	Taking the last two examples as a reference, we may want to 
	delete \textit{function specializations} based on their occurrence in the code.
	Suppose we have been using the \mycsnip{__attribute__}-based technique sketched in the last paragraph, and we wish to remove code specializations corresponding to the AVX512 ISA extension and its predecessor AVX2.
	Then a semantic patch for \textit{cleanup} could proceed as follows.
\begin{lstlisting}[language=cocci]
@c@
type T;
function f;
parameter list PL;
@@
- __attribute__((target(
(
- "avx512"
|
- "avx2"
)
- )))
- T f(PL) { ... }
@d@
type c.T;
function c.f;
parameter list c.PL;
@@
- __attribute__((target("default")))
  T f(PL) { ... }
\end{lstlisting}
	Two logical steps are needed, corresponding to two rules; here, \coccihelem{c} and \coccihelem{d}.
	Rule \coccihelem{c} removes all functions that are specific to one of the two mentioned instruction sets,
	identifiable by 
	being prefixed by a \mycsnip{__attribute__((target...}.
	Rule \coccihelem{d} merely removes the
	\mycsnip{__attribute__((target("default"))}
	specification, which precedes the unspecialized function definition, which is not deleted (in
	practice one would want to remove this attribute only once no specialization is there anymore).
	Rule \coccihelem{c} consists of a \textit{disjunction} matching the different
	attribute values with branches enclosed by parentheses and
        separated by \mycoccisnip{|} (all in the first column, to avoid confusion
        with the C/C++ bit-or syntax).
	Rule \coccihelem{d} reuses entities inherited from rule \coccihelem{c}, via the \textit{metavariables}
        \mycoccisnip{c.T}, \mycoccisnip{c.f}, and \mycoccisnip{c.PL}.
	Many bloat removal interventions are possible.
	Of the most pervasive and systematic ones, we imagine the location
        and removal of code associated with specific attributes or compiler-specific pragmas.

\para{Removal of explicit loop unrolling}

	We might obtain a codebase that we are unfamiliar with, containing thousands lines of code generated by a script, with loops unrolled, but no access to the original generator.
	Removal of many such explicit loops (or conceptually similar transformations) can be tricky, so here we show how we may approach this with Coccinelle.
	Let us assume that we want to target loops unrolled four times.
        More precisely, loops containing the same variation of a statement repeated four times,
        and each time with a different indexing of a specific subexpression,
        without local declarations, and most importantly, with \textit{the statement unknown}.
	The first semantic patch we show here consists of a rule named \coccihelem{p0}
	that locates four statements (\mycoccisnip{A}, \mycoccisnip{B}, \mycoccisnip{C}, \mycoccisnip{D}) in a loop (lines 14--22) with a specific loop header (lines 8--13) and checks that, given a loop variable \textit{matched by} metavariable \texttt{i}, these \textit{contain} respectively \texttt{i+0}, \texttt{i+1}, ... (lines 14, 16, 18, 20).
	Such a rule can match many simply unrolled loops, and with this arrangement of statements, removing \nobreakseq{\mycoccisnip{B},} \mycoccisnip{C}, and \mycoccisnip{D} removes any unrolling.

\begin{lstlisting}[language=cocci]
@p0@
type T;
identifier i,l;
constant k={4};
statement A,B,C,D;
@@
+ #pragma omp unroll partial (4)
  for (T i=0; i
-              +k-1
                    < l ;
-                         i+=k
+                         ++i
                              )
{
  \( A \& i+0 \) \(
- B \& i+1
  \) \(
- C \& i+2
  \) \(
- D \& i+3
  \)
}
\end{lstlisting}

	In \coccihelem{p0} we introduce
	the \textcolor{blue}{\texttt{\#pragma omp unroll partial}} clause,
	available since \OPENMP\ 5.1, which requests that the compiler
	enact loop unrolling, avoiding the need to modify the source code.

\begin{lstlisting}[language=cocci]
@p1@
type T;
identifier i,l;
constant k={4};
statement A,B,C,D;
@@
for (T i=0; i+k-1 < l; i+=k)
{
  \( A \& i+0 \) \( B \&
-    i+1
+    i+0
  \) \( C \&
-    i+2
+    i+0
  \) \( D \&
-    i+3
+    i+0
  \)
}

@r1@
type T;
identifier i,l;
constant k={4};
statement p1.A;
@@
+ #pragma omp unroll partial (4)
  for (T i=0; i
-              +k-1
                    < l ;
-                         i+=k
+                         ++i
                              )
{
  A
- A A A
}
\end{lstlisting}

In absence of further constraints, any sequence of four statements respectively
	accessing \texttt{i+0}, \texttt{i+1}... may be matched
	by \coccihelem{p0}, even if the sequence does not represent an
	unrolled loop, {\em i.e.}, the statements are not identical
	modulo \texttt{i+0}, \texttt{i+1}, etc..
	In certain codes this matching ambiguity may constitute a problem; for these cases
	rules \coccihelem{p1} and \coccihelem{r1} improve over \coccihelem{p0}.

	Just like rule \coccihelem{p0}, \coccihelem{p1} may also match more
 than we are interested into, because there is no guarantee that the
 statements are equal after substitution of the \mycsnip{i+0}... subexpressions.
	But the construction of \coccihelem{p1} is different:
	after locating four statements within a loop, on lines (10--11,
 13--14, 16--17), it replaces the
 occurrences of \mycsnip{i+0}, \mycsnip{i+1}... by \mycsnip{i+0} with the
 goal of making the four statements identical.
	The second rule \coccihelem{r1} will only match if all the substitutions of \coccihelem{p1} are enacted \textit{and} if they lead to a repetition of the first statement \nobreakseq{\mycoccisnip{A}.}
	That can only occur if the statements differed by the indexing expression only.
	In that case, \coccihelem{p1} will delete the last three statements \mycoccisnip{A}, and only keep the first one (the one on \mycsnip{i+0}).

If \coccihelem{p1} patches and transforms the code, but the resulting
statements are not identical, then \coccihelem{r1} will not match and the
loop will retain the four modified statements in the loop body, resulting in
code that is incorrect.  To address this issue, we could introduce a
third rule that undoes the transformations of \coccihelem{p1}
when \coccihelem{r1} is not applied.
	Once again, we showed a transformation that requires \textit{some}
        knowledge of the code, namely on its conventions (whether variable declarations are in the unrolled loop or outside, etc.).
	The code may also need some \textit{preparatory} modifications, enforcing certain conventions, before large-scale changes can be performed.

\para{Advanced expression modification (e.g., \texttt{mdspan})}
	Typically, a data layout change in a number-crunching codebase
	involves updating a large number of expressions accessing numeric arrays.
       	The following rule \coccihelem{tomultiindex} replaces triple nested square bracket expressions by \textsc{C++23}'s triple index expressions.
\begin{lstlisting}[language=cocci]
# spatch --c++=23
@tomultiindex@
symbol a;
expression x,y,z;
@@
- a[x][y][z]
+ a[x, y, z]
\end{lstlisting}%
This rule focuses on an array named \mycoccisnip{a}, but does not introduce any
requirement on the indexing expressions
	\mycoccisnip{x}, \mycoccisnip{y}, and \mycoccisnip{z}, which can be arbitrarily complex.
	Modifications of this sort are the most pervasive -- they usually aim at the entire code -- \textbf{ and probably are also of most interest to the readers} of this article.
	Of course, in order to be robust in production, the name of the redeclared arrays (here a hardcoded \mycoccisnip{a}) should probably follow a match in a global declaration, or perhaps a function-level parameter definition.

\para{Translation of very similar APIs}
	Some APIs may be so similar to each other, that their mutual translation would mostly consist of 
        token-to-token correspondence among two enumerable sets.
	This is the case of \textsc{Nvidia}'s CUDA \cite{NvidiaCUDA} and \textsc{AMD}'s HIP \cite{AmdHIP}, for which specific translation tools exists (e.g.,
	\texttt{hipify-perl}, based on \textsc{Perl}, and
	\texttt{hipify-clang}, based on \textsc{LLVM}).
	This sort of translation is also easy by means of Coccinelle. 
A Coccinelle \textit{toy} semantic patch for API translation, as illustrated below, could use a
\textsc{Python} rule\footnote{Coccinelle also provides an interface to \textsc{OCaml}.} to declare a dictionary (defined in the first, anonymous, special rule starting on line 1) to store the mappings of functions (this is exactly how \texttt{hipify-perl} does it, albeit without using an AST).

\begin{lstlisting}[language=cocci]
@initialize:python@ @@
C2HF = { "curand_uniform_double":
         "rocrand_uniform_double" }

@cfe@
identifier fn;
expression list el;
position p;
@@
 fn@p(el)

@script:python cf2hf@
fn << cfe.fn;
nf;
@@
coccinelle.nf =
  cocci.make_ident(C2HF[fn]);

@hfe@
identifier cfe.fn;
identifier cf2hf.nf;
position cfe.p;
@@
- fn@p
+ nf
  (...)
\end{lstlisting}

To translate the types as well, we proceed similarly,
using a rule \coccihelem{cte} to identify functions;
a rule \coccihelem{ct2hf} to use the dictionary;
and a rule \coccihelem{hte} to enact the change.

\begin{lstlisting}[language=cocci]
@initialize:python@ @@
C2HT = { "__half": "rocblas_half" }
@cte@
type c_t;
identifier i;
@@
 c_t i;

@script:python ct2hf@
c_t << cte.c_t;
h_t;
@@
coccinelle.h_t = \
  cocci.make_type(C2HT[c_t]);

@hte@
type ct2hf.h_t;
type cte.c_t;
identifier cte.i;
@@
- c_t i;
+ h_t i;
\end{lstlisting}

For simplicity, here we translate only one type and one function.
Rule \coccihelem{cfe} identifies CUDA function calls;
rule \coccihelem{cf2hf} uses the dictionary with the function translations to HIP;
finally, rule \coccihelem{hfe} enacts the change.

A complete version of these function- and type-level semantic patches would need to have the entire list
of functions and types involved in the two APIs.
That would be the same as the approach of \texttt{hipify-perl}, but with the advantage of
enactment of the changes at the AST level.

Among the further requirements of CUDA to HIP translation (which Coccinelle fullfills) is the ability to recognize the
\textit{triple chevron} syntax (\mycsnip{<<<>>>}) used by CUDA to launch GPU kernels,
and replace it with corresponding HIP library function calls, as HIP does not use the \textit{triple chevron}.

\begin{lstlisting}[language=cocci]
#spatch --c++
@@
identifier k;
expression b,t,x,y;
expression list el;
@@
- k<<<b,t,x,y>>>(el)
+ hipLaunchKernelGGL(k,b,t,x,y,el)
\end{lstlisting}

The rule above translates the invocation of a user-specified CUDA kernel (\mycoccisnip{k}) into HIP notation, which passes it to \mycsnip{hipLaunchKernelGGL} instead, along with all the existing arguments.
Notice that the rule is quite general, as it applies at the expression level 
and not at a statement level (no per-line replacement).
Translation to other
\mycsnip{hip}-prefixed kernel functions is to be addressed with trivial
separate rules.

\para{Translation of directive-based APIs}
	For very simple programs, translation of e.g., \OPENACC{} \cite{OACC} into \OPENMP{}
	may proceed on a line-by-line basis (a \texttt{\#pragma} directive occupies at least one line at a time)
	and without depending on the context.
	Moreover, the majority of projects adhere to using a specific subset of
	such directive-based APIs.
	So an approach of translation on a line basis may apply to a large
	number of cases.
	A skeleton semantic patch for such an approach follows:
\begin{lstlisting}[language=cocci]
@moa@
pragmainfo pi;
@@
  #pragma acc pi

@script:python o2o@
pi << moa.pi;
po;
@@
// Here we could have a small parser and translator using pi, but for simplicity we are just returning a hardcoded clause 
coccinelle.po =
  cocci.make_pragmainfo
    ("kernels copy(a)");

@@
pragmainfo moa.pi;
pragmainfo o2o.po;
@@
- #pragma acc pi
+ #pragma omp po
\end{lstlisting}
	In the above, the middle rule (\coccihelem{o2o}) invokes \textsc{Python} code.
	Please notice that in this example no variable initialization by means of \coccihelem{initialize:python} is needed.
	Such a \textsc{Python} rule could invoke a line-oriented parser-based translator implemented in place or in a separate \textsc{Python} module.
	In extracting the pragma lines, Coccinelle is liberal in accepting 
        whitespaces, and it does not break on line continuations, as an ad-hoc line-oriented script may do, so the parser would receive correct input.
	In addition to the individual \textcolor{blue}{\tt{\#pragma}} lines,
	more accurate translations would also require analyzing the context: that is also loops and local declarations.
	We are investigating such a translation-based approach for \OPENACC-to-\OPENMP\ in a student project.
	More complex uses of \textsc{Python} libraries would require
        stateful rule scripts, marked as such
	via the \coccihelem{initialize} and \coccihelem{finalize} keywords.

	The logic of translation could be not much different than that
        followed by the script provided by
	\textsc{Intel}
\footnote{\url{https://github.com/intel/intel-application-migration-tool-for-openacc-to-openmp/}}
	to assist with similar tasks, although the script provided by
	\textsc{Intel} lacks a proper parser or AST representation.

\para{Ease introduction of modern C++ STL constructs}
	C++ is a huge language and it allows vastly different styles of programming.
	For consistency reasons, most projects restrict the programmer to the use a subset of its features.
	Recently, the C++ committee has also been involved in suggesting programming guidelines \cite{CCG}.
	These may help avoid so-called \textit{code smells}, keeping code malleable and less error-prone.
	The following example replaces a so-called \textit{raw loop} with an STL-backed one
	(\mycsnip{std::find}).
	In a loop like this, apart from setting a flag
	(\mycsnip{result = true}), we could only perhaps expect some diagnostics (line 12), which in this example
	we delete.
\begin{lstlisting}[language=cocci]
#spatch --c++=17
@rl@
type T;
constant k;
identifier elem,result,arrid;
@@
-  bool result = false;
   ...
-  for ( T &elem : arrid )
-    if ( \( elem == k \| k == elem \) )
-      {
-        ...
-        result = true;
-        break;
-      }
+ const bool result =
+   (find(begin(arrid),end(arrid),k) !=
+    end(arrid));

@ah depends on rl@
@@
  #include <iostream>
+ #include <algorithm>
+ #include <functional>
\end{lstlisting}
	While not being particularly HPC-specific, this example shows that if there is sufficient regularity in our code, we may match specific recurring code portions and replace them by function calls (\mycsnip{find}).
	That is exactly what HPC-oriented C++ APIs usually require.
	The semantic patch we have shown also illustrates how intra-function declarations and constructs can be matched and manipulated.
	Notice how the \coccihelem{depends on} clause on the second rule 
	\coccihelem{ah} makes it apply conditionally on the first rule \coccihelem{rl}.

\para{Introduction of APIs enclosing lambdas}
Certain APIs
(e.g.: \textsc{Kokkos} \cite{edwards_trott_sunderland_2014_kokkos_data_layout_change},
\textsc{Raja} \cite{beckingsale_al_2019_raja}, 
ISO C++ standard parallelism \cite{CPP23}, and SYCL \cite{SYCL}) require enclosing the numerical kernels within
C++ \textit{lambda functions}, thus introducing one abstraction layer masking the exact parallel execution model on the hardware,
and allowing selection via e.g. definitions or template variables just before the lambda is instantiated.

\begin{lstlisting}[language=cocci]
#spatch --c++
@r0@ @@
+ #include <Kokkos_Core.hpp>
  #include <cmath>

@r1@
statement fb, fc;
expression n;
identifier c = {i,j};
position p;
@@
(
  fc@p
&
  for (...;c<n;...) fb
)

@script: python r2@
fb << r1.fb;
lb;
rp;
@@
coccinelle.lb =
 "KOKKOS_LAMBDA(const int i)" + fb;
coccinelle.rp =
 "RangePolicy<HostExecutionSpace>(0,n)";

@r3@
statement r1.fc;
position r1.p;
identifier r2.lb;
identifier r2.rp;
@@
(
  fc@p
&
(
- for (...;...;...) { ... result += ...; }
+ parallel_reduce(rp, lb);
|
- for (...;...;...) { ... }
+ parallel_for(rp, lb);
)
)
\end{lstlisting}

As of the most recent release of Coccinelle (version 1.3), lambda manipulations are not yet fully supported.
We nevertheless illustrate a loophole in Coccinelle that can be used to accomplish
certain transformations until the missing syntax becomes supported.

The above semantic patch illustrates this technique
being applied to a specific exercise
\footnote{\texttt{Exercises/01/Begin/exercise\_1\_begin.cpp} from \url{https://github.com/kokkos/kokkos-tutorials}, revision \texttt{fd4852c}.}
from the \textsc{Kokkos} tutorial.
Rule \coccihelem{r0} inserts the \textsc{Kokkos} header next to a header we may already include.
Rule \coccihelem{r1} locates \mycsnip{for} loops to be transformed into \textsc{Kokkos} loops;
after inspecting the source code associated to that exercise we established that the relevant loops
are exactly those with index variables \coccipelem{i} and \coccipelem{j}
(other loops exist, but we wish to keep them as they are).

In identifying the loops in this program we are content to use any \coccimelem{expression} \coccipelem{n} as loop bound.
We also declare two \coccimelem{statement} metavariables; one (\coccipelem{fc}) for the entire loop construct, and the other (\coccipelem{fb}) for the loop body.  Rule \coccihelem{r1} enacts no change: it only matches entities for reuse in the rule following it.
Rule \coccihelem{r2}, which uses \textsc{Python} scripting, makes an
anonymous lambda function out of the entire block \coccipelem{fb}
representing the body of the \mycsnip{for(...)} loop, and saves that in a \textsc{Python} string.
Rule \coccihelem{r3} matches different possible loop bodies to recognize the different \textsc{Kokkos} parallelization constructs they belong to (normal parallelism vs.\ reduction).
Rule \coccihelem{r3} exploits the aforementioned intentional loophole in Coccinelle,
which
allows treating any string as an \coccimelem{identifier} metavariable, and
passes the string in calls to \textsc{Kokkos}' \mycsnip{parallel_for}
and \mycsnip{parallel_reduce}, in a context that would otherwise accept
identifiers.

This \textit{dirty hack} of passing statements through \textsc{SmPL} identifiers allows a limited manipulation of lambdas with Coccinelle.
We plan to implement proper lambda manipulation support soon.

\para{Workarounds for occasional compiler bugs}
A few years ago\footnote{Private communication by the
Debian-side maintainer to the first author, who is also author of \textsc{LIBRSB}.}, 
failure of multiple checks was reported in the test suite
of the \textsc{librsb} library, which led to the suspicion of a bug there \cite{librsb}.
The failing involved double precision complex conjugate computations.
A brief investigation showed that the bug was in the \textit{vectorizer} of
the GCC
compiler,\footnote{\url{https://gcc.gnu.org/bugzilla/show_bug.cgi?id=103995}}
version 11.2 (and a few other ones after).
Considering that the library was being packaged on several \textsc{Linux}
distributions, the bug needed a portable fix that would possibly only be activated when using the affected compiler versions.
The decision of the \textsc{LIBRSB} author was to modify the build system (\texttt{Makefile}s \& co.) to trigger a semantic patch-based transformation, conditional on the compiler version.
This semantic patch applies only to the functions that have been identified as being affected by the compiler bug, and adds specific \mycsnip{#pragma} lines lowering the optimization level, so to get correct numerical results.
A regular expression can be used to identify the affected functions because the code uses a specific naming convention for the functions doing double precision complex conjugation.
\begin{lstlisting}[language=cocci]
@pragma_inject@
identifier i =~ "rsb__BCSR_spmv_sasa_double_complex_[CH]__t[NTC]_r1_c1_uu_s[HS]_dE_uG";
type T;
@@
+ #pragma GCC push_options
+ #pragma GCC optimize "-O3", "-fno-tree-loop-vectorize"
T i(...)
{
 ...
}
+ #pragma GCC pop_options
\end{lstlisting}
The semantic patch impacts a dozen functions among a few hundred; its attractiveness derives mostly from its inobtrusiveness and for being triggerable on demand.
A similar approach could be used by other projects to
implement \textit{transitory changes} conditionally via \textit{transitory semantic patches}.

\section{Discussion}
\label{sec:discu}
\para{Acceptance obstacles}
Code owners or maintainters may resist large-scale modifications of the
codebase, especially if such restructurings originate from a different team.
The habit of writing comprehensive test suites in the scientific software
communities is slowly but steadily progressing;
this can surely facilitate reviewing a large ``refactoring contribution'' on the basis of the tests' results, notwithstanding that even the best test suite cannot be considered complete.
Apart from simple mistrust, other factors hindering acceptance can be aesthetic (with some relation to intellegibility) or doubts about the maintainability of the new code.
We believe that the spread and adoption of sound \textit{research software engineering} practices can reduce acceptance impediments while raising confidence in the code behaviour and its malleability.

\para{Replayable refactorings}
Coccinelle can enable developing restructurings over time by making it
possible to keep an existing code separate from the change specifications.
That is, as an existing code base can evolve, and semantic patches can be developed in parallel.
This gives the ability to apply a refactoring on demand, and not having to maintain overly similar branches in parallel.
Once the \textit{refactoring specification} is thoroughly validated (apart from tests,
that may include inspection and study), the authors may be willing to accept the resulting patched version.
Note that the large-scale modifications may also impact the results of tests.
For instance, parameters order may change, or involved variables' types.
In that case we would want to update the unit test source files accordingly as well.

\para{The future of C/C++}
Recent trends in programming languages and organizational 
security discourage the use of C/C++, in favour of intrinsically safer languages.
Given the sheer size of the installed user base, any migration
to a new safer programming language will take a long time.
During that time, tooling for programming responsibly,
to mitigate existing risks or weaknesses,
and perhaps migrating to other languages, will continue
to be needed. 

Even if exhibiting a lower urgency than in other categories of software
(e.g., medical, avionics), HPC software deserves correctness and
reliability.  If such code is ever ported to or developed in \textsc{Rust}, the
resulting code may still be able to benefit from the use of Coccinelle, as
a \textsc{Rust} port of Coccinelle is under
way.\footnote{\url{https://rust-for-linux.com/coccinelle-for-rust}}
 
And while Coccinelle is not a tool for translating \textit{between}
languages, it is possible to conceive an iterative \textit{migration} process
consisting of partial rewrites of program modules or classes,
and subsequent replacement of old constructs with new ones referring
to new functions and classes in the different language.
The viability of such an approach is clearly highly project-dependent.

Whatever the programming model or next best \textit{performance portability library} of interest may be,
our work aims at aiding otherwise unfeasible or work-intensive, large-scale code
changes in a possibly reversible (or \textit{replayable}) manner.

\section{Conclusion}
\label{sec:conc}
We have succinctly described a number of use cases
illustrating the construction of the Coccinelle semantic patch language
which can be of interest to HPC practitioners.

We like to regard such cases as individual ``components'' in typically larger, more involved HPC refactorings.
These refactorings are primarily aimed at
separating the expertise of machine-specific experts from that of the
domain-specific expert.

The techniques we show are more effective and productive when the codebase is
\textit{tidy} and regular.
In itself, such properties are desirable also when an expert
is refactoring directly with a code editor/IDE and basic text-editing tools.

A challenge in working with C++ is the large size and the continual, vigorous
evolution of the language.  Development of support for C++ code in
Coccinelle is ongoing.
\balance
\ifodd\WantIEEETran
\bibliographystyle{IEEEtran}
\else
\bibliographystyle{alpha}
\fi
\bibliography{hips25_paper_arxiv}

\newcommand{\etalchar}[1]{$^{#1}$}
\begin{thebibliography}{FBB{\etalchar{+}}99}

\bibitem[Amd]{AmdHIP}
{AMD HIP (Heterogeneous-Compute Interface for Portability)}.
\newblock \url{https://rocmdocs.amd.com/projects/HIP/}.

\bibitem[BIHK16]{BaruffaIHK16}
Fabio Baruffa, Luigi Iapichino, Nicolay~J. Hammer, and Vasileios Karakasis.
\newblock Performance optimisation of smoothed particle hydrodynamics
  algorithms for multi/many-core architectures.
\newblock {\em CoRR}, abs/1612.06090, 2016.

\bibitem[BSB{\etalchar{+}}19]{beckingsale_al_2019_raja}
David Beckingsale, Thomas R.~W. Scogland, Jason Burmark, Rich Hornung, Holger
  Jones, William Killian, Adam~J. Kunen, Olga Pearce, Peter Robinson, and
  Brian~S. Ryujin.
\newblock {RAJA:} portable performance for large-scale scientific applications.
\newblock In {\em 2019 {IEEE/ACM} International Workshop on Performance,
  Portability and Productivity in HPC, P3HPC@SC 2019, Denver, CO, USA, November
  22, 2019}, pages 71--81. {IEEE}, 2019.

\bibitem[C23]{C23}
The {C23} standard draft.
\newblock \url{https://www.open-std.org/jtc1/sc22/wg14/www/docs/n3096.pdf}.

\bibitem[CCG]{CCG}
{C++} core guidelines.
\newblock \url{https://isocpp.github.io/CppCoreGuidelines/CppCoreGuidelines}.

\bibitem[CPP]{CPP23}
The {C++23} standard draft.
\newblock
  \url{https://www.open-std.org/JTC1/SC22/WG21/docs/papers/2023/n4950.pdf}.

\bibitem[CRE]{CRE}
{CLANG}'s refactoring engine.
\newblock \url{https://clang.llvm.org/docs/RefactoringEngine.html}.

\bibitem[ETS14]{edwards_trott_sunderland_2014_kokkos_data_layout_change}
H.~Carter Edwards, Christian~R. Trott, and Daniel Sunderland.
\newblock Kokkos: Enabling manycore performance portability through polymorphic
  memory access patterns.
\newblock {\em J. Parallel Distributed Comput.}, 74(12):3202--3216, 2014.
\newblock \url{https://doi.org/10.1016/j.jpdc.2014.07.003}.

\bibitem[FBB{\etalchar{+}}99]{Fowler1999}
Martin Fowler, Kent Beck, John Brant, William Opdyke, and Don Roberts.
\newblock {\em Refactoring: Improving the Design of Existing Code}.
\newblock Addison-Wesley Professional, 1999.
\newblock {ISBN: 0-201-48567-2}.

\bibitem[GFM]{GFM}
{GCC}: Function multiversioning (attributes \texttt{target} and
  \texttt{target\_clones}).
\newblock
  \url{https://gcc.gnu.org/onlinedocs/gcc/Common-Function-Attributes.html}.

\bibitem[lib]{librsb}
{LIBRSB}: {Sparse BLAS based on the Recursive Sparse Blocks Format}.
\newblock \url{https://librsb.sf.net/}.

\bibitem[LM18]{lawall_muller_2018_ten_years_atc}
Julia Lawall and Gilles Muller.
\newblock Coccinelle: 10 years of automated evolution in the {Linux} kernel.
\newblock In {\em 2018 {USENIX} Annual Technical Conference, {USENIX} {ATC}},
  pages 601--614, 2018.

\bibitem[Mar]{martone_20250122_coccinelle_training}
Michele Martone.
\newblock {Introduction to Semantic Patching of C and C++ Programs with
  Coccinelle}.
\newblock \url{https://doi.org/10.5281/zenodo.14728519}.

\bibitem[ML21]{c3po}
Michele Martone and Julia Lawall.
\newblock Refactoring for performance with semantic patching: Case study with
  recipes.
\newblock In Heike Jagode, Hartwig Anzt, Hatem Ltaief, and Piotr Luszczek,
  editors, {\em High Performance Computing - {ISC} High Performance Digital
  2021 International Workshops, Frankfurt am Main, Germany, June 24 - July 2,
  2021, Revised Selected Papers}, volume 12761 of {\em Lecture Notes in
  Computer Science}, pages 226--232. Springer, 2021.

\bibitem[Nvi]{NvidiaCUDA}
{Nvidia CUDA C++ Programming Guide}.
\newblock \url{https://docs.nvidia.com/cuda/pdf/CUDA_C_Programming_Guide.pdf}.

\bibitem[OAC]{OACC}
{OpenACC 3.2 Specification}.
\newblock
  \url{https://www.openacc.org/sites/default/files/inline-images/Specification/OpenACC-3.2-final.pdf}.

\bibitem[OMP]{OMP6}
{OpenMP} 6.0 {Specification}.
\newblock
  \url{https://www.openmp.org/wp-content/uploads/OpenMP-API-Specification-6-0.pdf}.

\bibitem[PHS15]{Pennycook2015143}
Simon~J. Pennycook, Christopher~J. Hughes, and Mikhail Smelyanskiy.
\newblock Chapter 8 - optimizing gather/scatter patterns.
\newblock In James Reinders and Jim Jeffers, editors, {\em High Performance
  Parallelism Pearls}, pages 143 -- 157. Morgan Kaufmann, Boston, 2015.

\bibitem[QL11]{quinlan_2011_rose}
Dan Quinlan and Chunhua Liao.
\newblock The {ROSE} source-to-source compiler infrastructure.
\newblock In {\em Cetus users and compiler infrastructure workshop, in
  conjunction with PACT}, volume 2011, page~1, 2011.

\bibitem[ros]{rosewww}
The {ROSE} compiler webpage.
\newblock \url{https://rosecompiler.org/}.

\bibitem[{Spr}05]{springel2005cosmological}
V.~{Springel}.
\newblock {The cosmological simulation code GADGET-2}.
\newblock {\em MNRAS}, 364:1105--1134, 2005.

\bibitem[SYC]{SYCL}
{SYCL: C++ Programming for Heterogeneous Parallel Computing}.
\newblock \url{https://www.khronos.org/sycl/}.

\end{thebibliography}
\end{document}